\PassOptionsToPackage{svgnames,table}{xcolor}

\documentclass[preprint]{vgtc}               

\graphicspath{{figures/}{pictures/}{images/}{./}} 

\usepackage{times}                     

\usepackage{tabu}                      
\usepackage{booktabs}                  
\usepackage{lipsum}                    
\usepackage{mwe}                       
\usepackage{amsmath}
\usepackage{multirow}
\usepackage{mathptmx}                  
\usepackage[table]{xcolor}
\usepackage{makecell}
\usepackage{graphicx}
\usepackage{tikz}
\usepackage{tabularx}
\usepackage[export]{adjustbox}

\definecolor{sparkBlueLight}{RGB}{135, 185, 215}  
\definecolor{sparkGreenLight}{RGB}{145, 205, 145} 
\definecolor{sparkOrangeLight}{RGB}{245, 130, 30} 
\definecolor{sparkBlue}{RGB}{0, 51, 255}  
\definecolor{sparkOrange}{RGB}{255, 64, 0} 
\definecolor{sparkGreen}{RGB}{0, 158, 96} 
\definecolor{sparkSep}{RGB}{200, 200, 200}   

\newcommand{\growthsparkbezierarrow}[1]{%
    \begin{tikzpicture}[
        baseline={(0,0.0em)}, 
        x=0.66em, y=0.66em, 
        line cap=round, 
        line join=round
    ]

        \def\myColor{sparkGreen}
        \def\L{#1} 

        \fill[\myColor!40]
            (0,0) -- (0,0) 
            [rounded corners=0.07em] to[out=2, in=195] ({1.2*\L}, 0.4) 
            to[out=15, in=200] ({2.4*\L}, 1.2) -- ({2.4*\L}, 0) -- cycle;

        \draw[\myColor, line width=1.1pt] 
            (0,0) [rounded corners=0.07em] 
            to[out=2, in=195] ({1.2*\L}, 0.4) 
            to[out=15, in=200] ({2.4*\L}, 1.2);

        \draw[
            \myColor, 
            line width=1.2pt, 
            {Stealth[scale=0.45]}-{Stealth[scale=0.45]}
        ] ({2.4*\L + 0.4}, 0.05) -- ({2.4*\L + 0.4}, 1.35); 
    \end{tikzpicture}%
}

\usepackage{tikz}
\usepackage{xspace}
\usepackage{graphicx}
\usepackage{xcolor}
\usetikzlibrary{shadings, arrows.meta}

\newcommand{\blueC}{%
  \tikz[baseline=-0.7ex] \fill[blue] (0,0) circle (0.5em);%
}

\newcommand{\numberedpdf}[2]{%
  \begin{tikzpicture}[baseline=(image.center)]
    \node[inner sep=0pt] (image) {%
      \includegraphics[width=\linewidth]{#2}%
    };
    \node[
      anchor=north west,
      xshift=2mm,
      yshift=-2mm,
      inner sep=0pt,
      font=\bfseries\color{gray}
    ] at (image.north west) {#1};
  \end{tikzpicture}%
}

\definecolor{wpmblue}{HTML}{2166AC}
\definecolor{wpmred}{HTML}{B2182B}
\definecolor{wpmgray}{HTML}{F3F4F6}

\definecolor{ratingLow}{HTML}{D55E5E}
\definecolor{ratingHigh}{HTML}{4A78C2}

\newcommand{\lowcell}[2]{%
    \cellcolor{ratingLow!#1!white}#2%
}
\newcommand{\highcell}[2]{%
    \cellcolor{ratingHigh!#1!white}#2%
}
\newcommand{\midcell}[1]{%
    \cellcolor{white}#1%
}
\onlineid{7265}

\vgtccategory{Research}

\vgtcinsertpkg

\newcommand{\hsw}[1]{{\color{blue}[HSW: #1]}}
\newcommand{\cx}[1]{\textcolor[RGB]{232, 125, 114}{[CX: #1]}}
\newcommand{\marti}[1]{{\color{purple}[MH: #1]}}

\newcommand{\revadd}[1]{{\color{black}#1}}
\newcommand{\revdel}[1]{{\color{red}#1}}
\renewcommand{\revdel}[1]{}

\title{Visual Embellishments are Potential Distractions\\ in Double-Column Reading}

\author{Songwen Hu\thanks{e-mail: shu343@gatech.edu}\\ %
        \scriptsize Georgia Institute of Technology %
\and Chase Stokes\thanks{e-mail: cstokes@ischool.berkeley.edu}\\ %
     \scriptsize University of California, Berkeley %
\and Marti A. Hearst\thanks{e-mail: hearst@berkeley.edu}\\ %
     \scriptsize University of California, Berkeley %
\and Cindy Xiong Bearfield\thanks{e-mail: cxiong@gatech.edu}\\ %
     \scriptsize Georgia Institute of Technology}
     
\teaser{
\vspace{-1.5em}
    \centering
    \setlength{\tabcolsep}{4pt}
    \renewcommand{\arraystretch}{1.2}

    \begin{tabular}{@{}>{\centering\arraybackslash}m{1cm}
                    >{\centering\arraybackslash}m{5cm}
                    >{\centering\arraybackslash}m{5cm}@{}}
        & \textbf{Not Embellished} & \textbf{Embellished} \\

    \makebox[\linewidth][c]{\rotatebox[origin=c]{90}{\textbf{Data-heavy}}} &
    \numberedpdf{1}{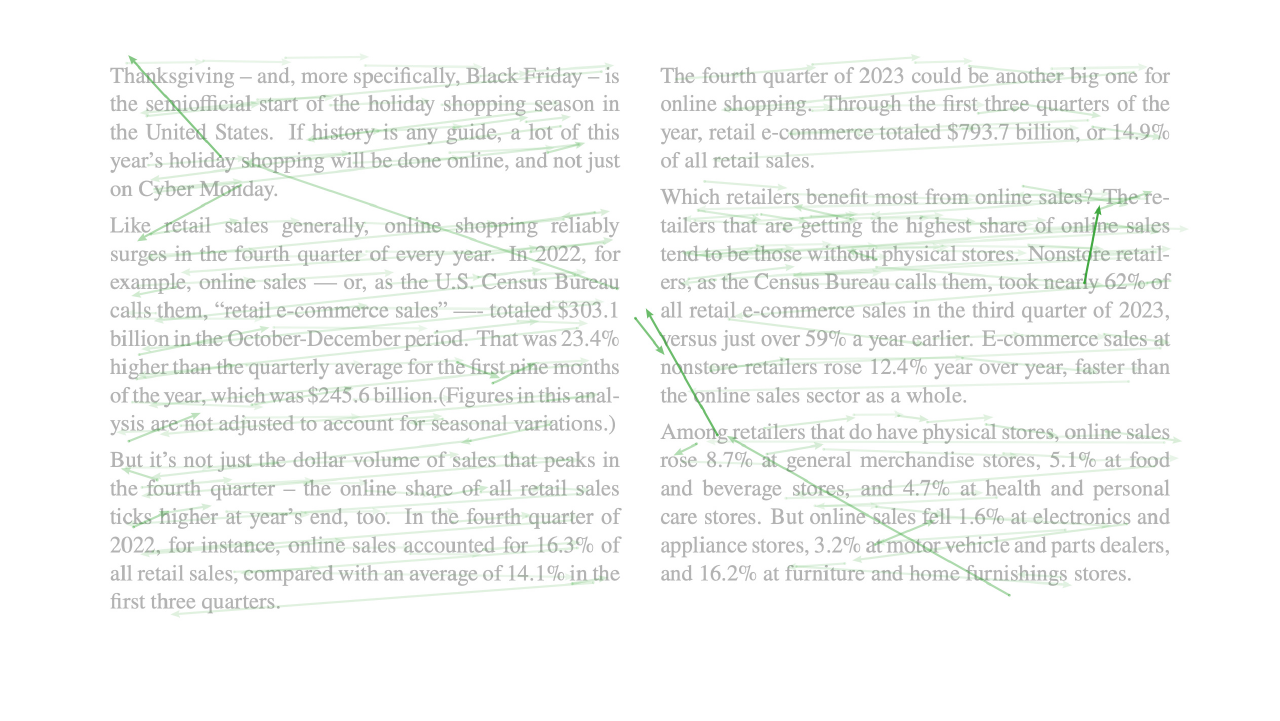} &
    \numberedpdf{3}{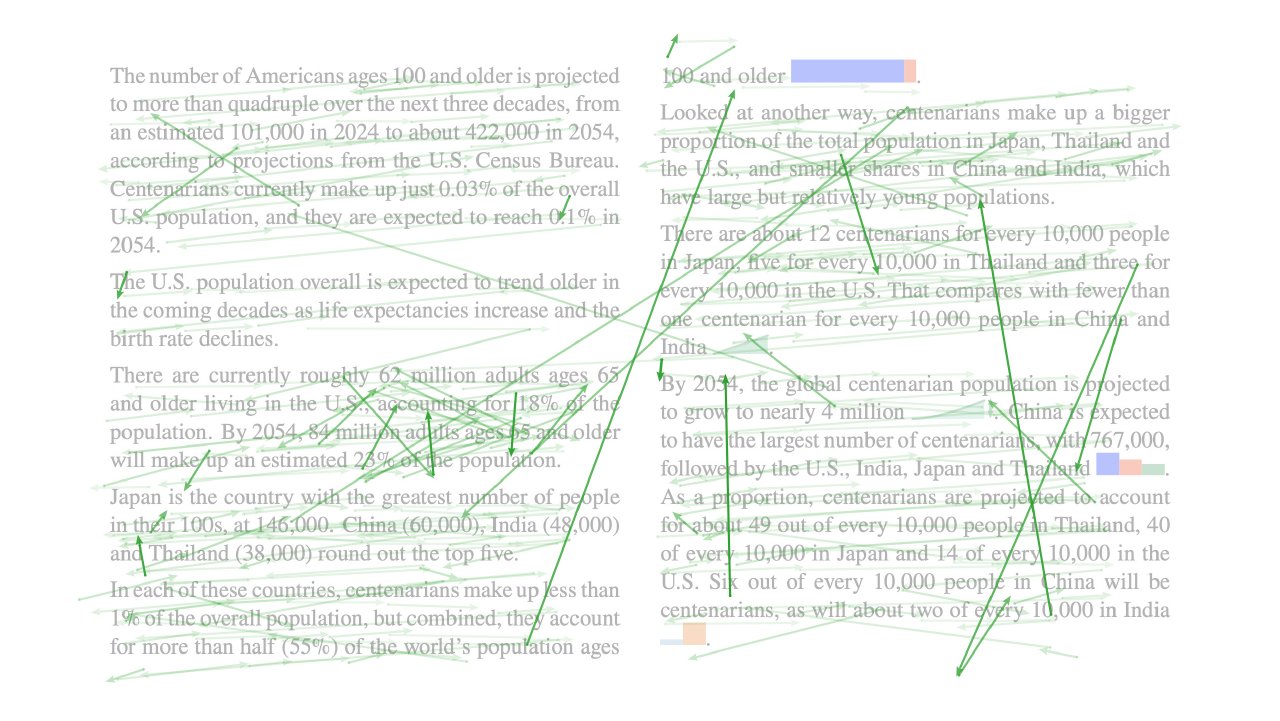} \\
    
   \makebox[\linewidth][c]{\rotatebox[origin=c]{90}{\textbf{Data-light}}} &
    \numberedpdf{2}{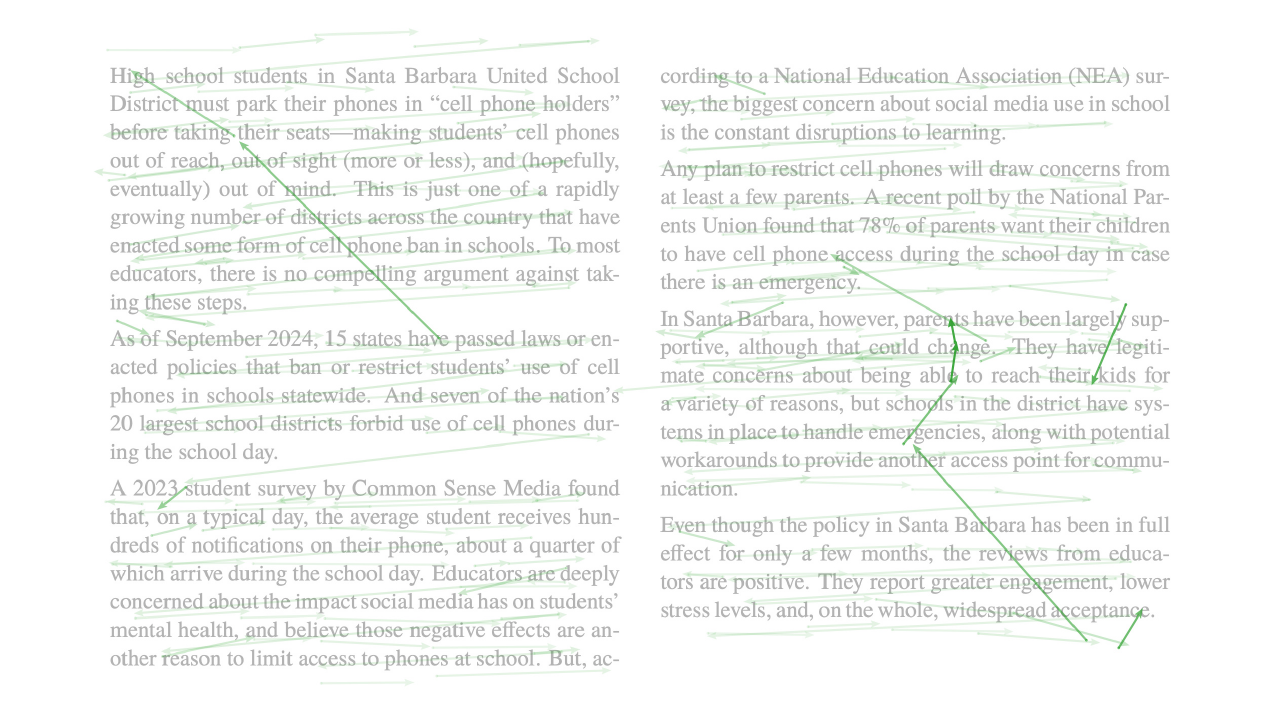} &
    \numberedpdf{4}{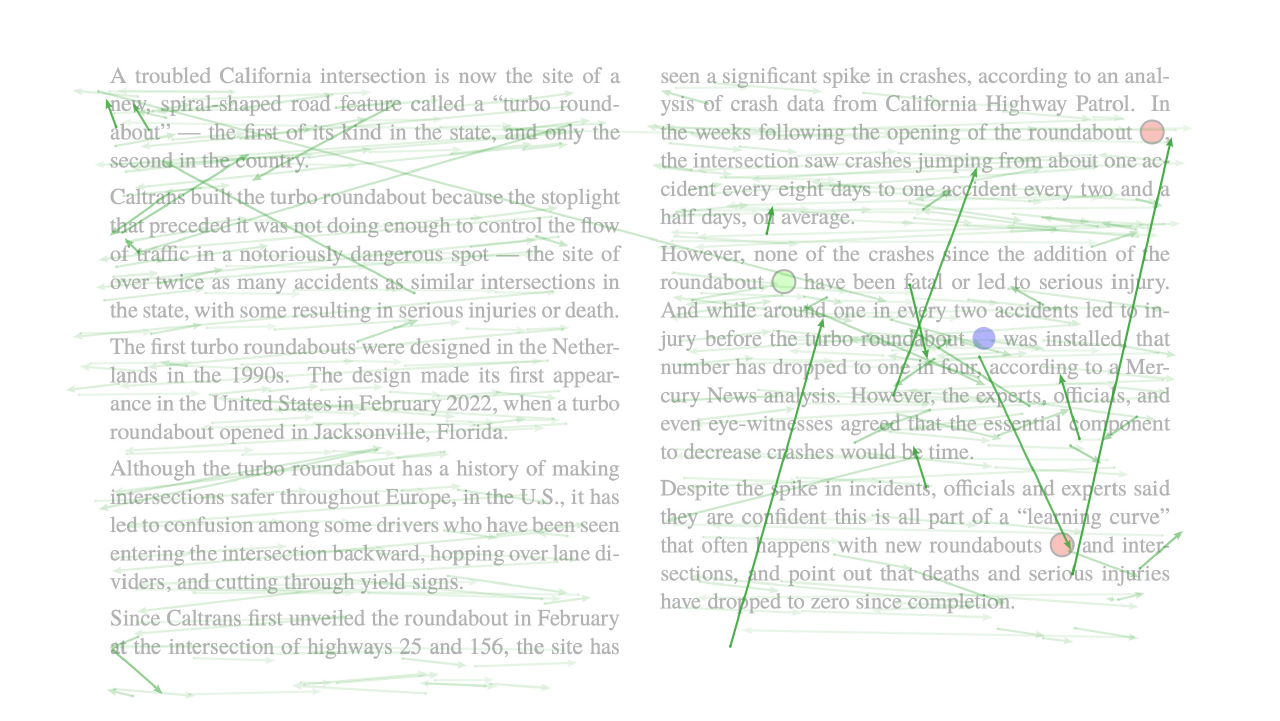}
    \end{tabular}
\vspace{-1.5em}
    \caption{Participant P2's gaze path pattern across data density
    and visual embellishment conditions. 
    Saccade opacity is encoded according to the deviation of the saccade direction from the horizontal, with larger deviations shown at higher opacity. 
    \revadd{For this participant, } both numerical values and visual embellishments are associated with noisier saccades than non-embellished conditions.
    Numbers in the upper left of each trial indicate randomized presentation order for this participant.  
    }
    \label{fig:p2-condition-comparison}

}

\abstract{
Eye-catching graphics, such as circular figure labels and word-scale visualizations, are increasingly being placed directly within long-form text paragraphs. Some research has claimed that inline visualizations can help readers understand data-rich passages more clearly. However, research in the science of reading   calls into question the introduction of images within the flow of text.
In this work, we conduct an exploratory study of eye movement in both the presence and absence of visual embellishments.  Using a high-resolution eye-tracker (EyeLink Portable Duo), \revdel{a study with six participants finds more variation in eye movement with the presence of visual embellishment than without, without clear evidence of improved comprehension.  Furthermore, participants rated the visual embellishments to be  distracting compared to the plain text.  These results  conflict with prior findings, indicating that further study with larger sample sizes and more varied designs are needed to clarify the potential drawbacks of inline graphics.} \revadd{we observed small mean increases in vertical and cross-column saccade rates among six participants, with substantial variation among readers and no detected difference in comprehension accuracy. Mean subjective ratings for the data-light circular-glyph passage indicated it is more distracting than its unembellished comparison passage. These exploratory observations differ from prior findings and motivate a larger, fully crossed study of inline graphics.}
}

\CCScatlist{
  \CCScatTwelve{Human-centered computing}{Visu\-al\-iza\-tion}{}{Visualization design and evaluation methods}
}

\nocopyrightspace

\begin{document}


\maketitle

\section{Introduction}

Many observers are noting with distress the decline of long-form reading and reduced attention-span in the population at large \cite{naep2024,haidt2024anxious}.  This can be attributed to many factors, including split attention caused by short-form media, the proliferation of video content, and a decline in the habit of reading \cite{postman2005amusing,Mago2020BooksIT}.  Reading has been shown to promote deeper thinking and bolster the ability to reason and think logically and causally; deep, concentrated reading must be taught and takes practice to become effortless \cite{wolf2018reader}.  Successful reading is also bolstered by appropriate text presentation, requiring care  in text layout and contrast, among other factors \cite{Miniukovich2017DesignGF,Scaltritti2019InvestigatingEO}.

Digital media increasingly relies on dynamic visual strategies to capture and sustain people's attention. 
Recently, this trend toward highly stimulating visual presentation has permeated document design and academic publishing. Modern manuscripts frequently incorporate rich graphical abstracts, color-coded semantic highlights, and decorative visual embellishments embedded directly within or alongside the text.  Scholarly scientific documents are increasingly including paragraph-embedded images \cite{Lu2026GraphingIU,Hong2026ReviewAA}, and word-scale embellishments within text are prevalent in the information visualization community (e.g., \cite{beck2016expert,goffin2016exploratory,zou2025gistvis}).  

Studies of word-scale visualizations have claimed that these graphics improve comprehension by providing readers with multiple modes of encoding and processing information. However, such embellishments also run the danger of impeding or interfering with readability, especially for long or complex texts that should be read deeply.   They may also impede the ability of readers to scan such documents to gain an overview of their contents.  However, empirical studies often do not compare such visuals to alternatives, such as presenting visualizations in blocks between paragraphs so as to not interrupt the reading flow.  Studies also do not closely investigate if such designs impede the ability to read fluently.

In this work, we use a high-resolution eye tracker to investigate the effects of within-paragraph visual embellishments on eye movement. Based on the science of reading, we hypothesize that in-line visual embellishments interrupt the flow of reading without providing commensurate benefit. \revdel{We conduct a small exploratory study with six participants, finding both empirical and qualitative evidence that within-paragraph visual embeddings are distracting to the reading process.}\revadd{In a six-participant exploratory study, we observed small descriptive differences in gaze behavior and lower subjective ratings for one embellished condition. We use these observations to suggest hypotheses for larger studies.}

\section{Related Work}

Inspired by Tufte's sparklines \cite{tufte2006beautiful}, word-scale visualizations are text-sized graphics deployed to augment information   \cite{goffin2014exploring}.  Domain-specific contexts that have been explored include programming \cite{hoffswell2018augmenting}, sports statistics \cite{Perin2013SoccerStoriesAK},  multivariate data \cite{brandes2013gestaltlines},  and eye tracking data \cite{beck2016expert}. There is evidence that sparklines in tabular format can ease the cognitive
load of finding trends in large volumes of data. For instance, when
used for analyses of medical \cite{bauer2010design} or accounting \cite{parsons2013testing} tabular data,
sparklines allowed for faster interpretation and easier comparison of
patterns than tabular data alone. On the other hand, they sometimes
lack the precision needed for specific data lookup or identification of
anomalies, due to their small size, lack of vertical axes, and aspect
ratio\revadd{~\cite{parsons2013testing}}. 


\smallskip
\noindent\textbf{Word-scale visualizations placed into text:}
As highlighted in a  survey by Lan et al.~\cite{lan2026evolving}, the integration of textual and visual modalities has the potential to enhance the reading experience by facilitating a deeper, more immediate understanding of the data-rich documents without draining the readers' cognitive resources \cite{goffin2014exploring, beck2017word}.
However, empirical results are mixed as to the efficacy of embedding word-scale graphics directly within text paragraphs, with researchers noting their sensitivity to the context of use and size of the graphic.

On the side of effectiveness,
Goffin et al.~\cite{goffin2015reading} showed 12 CS PhD student participants visualizations in four positions beside isolated double-spaced sentences. The study included training materials about the meanings of the visualizations, and the test questions asked about the trends shown. No significant difference was found for reading time or error rate. Placing the visualization above the sentence was the most preferred position, but can be argued that this is akin to placing the image outside of the text. Three participants ranked no visualization as their second choice.

More recently, Zou et al.~\cite{zou2025gistvis} showed  12 participants interactive word-scale visualizations and text highlights. The study design
showed the passage next to the questions and included a training session about the meaning of the word-scale visualizations. While there were no differences found in time or accuracy from a plain text condition, the visualization conditions received better scores for perceived mental demand and effort. It is unclear how much this is attributable to the highlighted text versus the graphics, and some comments about the graphics were negative.  


In other contexts, embedding graphics directly within running text introduced reading delays due to the cognitive overhead of processing multiple modalities simultaneously within a single sentence.
In a study replacing a single word in a stand-alone sentence with a congruent emoji, Cohn et al. \cite{cohn2018emoji} discovered that while emojis increase overall reading time, they do not affect a participant's subjective perception of their own comprehension. Reading times also increased on the words immediately following the emoji, indicating a lingering cognitive disruption that propagates throughout the sentence \cite{cohn2018emoji}. 
When the emoji was instead placed adjacent to the target word rather than replacing it, Barach et al. \cite{barach2021emojis} observed a different processing dynamic: participants frequently skip over the target word entirely to prioritize the emoji, yet overall sentence reading times did not reduce, likely because attention was unevenly distributed between the text and the emoji. 

Huth et al. \cite{huth2024eye} used eye tracking to study double-spaced short texts (95-115 words) containing a combination of text highlighting, icons, and word-scale graphics\revdel{. found}\revadd{, finding} promising trends for reading support and memory, but also observing that readers looked at added highlights, icons, and word-sized visualizations frequently, potentially attention breaking flow and described as annoying by some participants. They conclude that glyphs should be used sparingly and only for key parts of the text.

\noindent
\textbf{Visual saliency and involuntary attentional capture:} 
Although word-scale visualizations were designed to eliminate split attention \cite{ayres2005split} by co-locating text and graphics \cite{beck2017word}, these inline elements can function as highly salient visual embellishments. Consequently, while the resulting reading delays might theoretically foster deeper reflection \cite{lin2026four}, existing work in human perception literature suggests they may instead capture attention involuntarily and degrade overall reading quality.

Reading spatially-separated figures and text requires top-down, \textit{voluntary} divided attention as readers deliberately coordinate focus between text and graphics~\cite{connor2004visual}. 
Conversely, word-scale visualizations can induce \textit{involuntary} divided attention through bottom-up attentional capture~\cite{theeuwes2025attentional}; these salient elements automatically pull spatial focus away from the primary foveal reading task regardless of user intent~\cite{carrasco2011visual, franconeri2003moving}.
They act as distractors that compromise target selection and weaken the visual system's capacity to isolate the primary target text \cite{ceja2023limits}. 
Ceja et al.~\cite{ceja2023limits} found that when peripheral objects successfully capture attention, the resulting visual noise actively distorts feature processing at the focal point of vision, leading to visual feature misbindings and perceptual clutter near the fovea. 
Filtering out these intrusive peripheral distractors often require active cognitive control, which drains limited working memory capacity and ultimately reduces higher-level comprehension, memory retention, and information-processing efficiency \cite{lavie2004load}. 
\revadd{Similar phenomena were also  observed in studies of high-level visualization comprehension \cite{quadri2024see, fygenson2025cognitive}.}

Salient visual distractors can trigger \emph{saccadic inhibition}, which interrupts eye-movement planned by the brain and prolongs reading fixations \cite{reingold2004saccadic}. 
However, the downstream cognitive consequences of such visual distraction remain underexplored, especially for reading scientific journals that require meticulous semantic processing.
Our paper presents a preliminary evaluation using eye-tracking to examine \textbf{whether and how word-scale visualizations disturb readers's reading}, while exploring their \textbf{downstream effects on comprehension and the perceived reading experience}.

\section{Study Design}
\label{sec:study-design}

We conducted a within-subject eye-tracking study to examine how compact visual embellishments embedded in two-column articles affect reading behavior. We tracked participants' eye movements, particularly ones that crossed between columns or otherwise departed from the predominantly horizontal progression of text reading. 
Each participant read four passages spanning two levels of data intensity (heavy and light), with or without visual embellishments. 

\begin{table}[htbp]
\centering
\setlength{\tabcolsep}{2pt}

\begin{adjustbox}{max width=\linewidth}
\begin{tabular}{llrll}
\toprule
 & \textbf{Passage Name}
 & \textbf{\# Words}
 & \textbf{Source}
 & \textbf{Embellishment} \\
\midrule
\textbf{\multirow{2}{*}{Data-heavy}}
 & \textit{Centenarians}
 & 364
 & \cite{centenarians} via \cite{zou2025gistvis}
 & Word-scale \\
 & \textit{Black Friday}
 & 342
 & \cite{blackfriday} via \cite{zou2025gistvis}
 & None \\
\midrule
\textbf{\multirow{2}{*}{Data-light}}
 & \textit{Roundabout}
 & 334
 & \cite{roundabout}
 & Circular Glyph \\
 & \textit{Smartphones}
 & 339
 & \cite{smartphones}
 & None \\
\bottomrule
\end{tabular}
\end{adjustbox}

\caption{Stimuli characteristics.}
\label{tab:stimuli}
\vspace{-2em}
\end{table}

\subsection{Stimuli Design}
\label{sec:stimuli}

As summarized in Table~\ref{tab:stimuli}, we adapted four articles of comparable length into a $2 \times 2$ experimental design crossing \textit{Data Density} (Data-Heavy vs. Data-Light) with \textit{Visual Embellishment} (Embellished vs. Unembellished Baseline).
Data-heavy passages contained dense numerical information, whereas data-light passages featured predominantly narrative prose. 
\revdel{To reflect real-world publishing practices,}\revadd{To cover two common embellishment forms observed in academic publishing,} the type of embellishment was tailored to the passage's data intensity: the data-heavy \textit{Centenarians} passage incorporated quantitative word-scale visualizations embedded within the text (modeled after \cite{zou2025gistvis}) \growthsparkbezierarrow{1.5} to redundantly encode the data, and is paired against the  \textit{Black Friday} baseline, which is data-heavy but unembellished; 
the data-light \textit{Roundabouts} passage featured inline colored circular glyphs placed adjacent to each mention of the word ``roundabout'' \blueC,
mirroring common trends in academic publications of embedding small icons and symbols next to a key word, and this is
paired against the \textit{Smartphones} baseline, which is data-light but unembellished.
This grouping allows comparison of embellishment use in both numerically dense and comparatively narrative text. Passages were modified and condensed to have a comparable length.


The passages were set in the commonly used double-column journal format, reflecting the design of academic papers.
The LaTeX \revdel{\texttt{tixz}}\revadd{\texttt{TikZ}} package was used to render all inline embellishments.
Two-column PDFs were generated with the following settings, in pixels: paper width: 2560, 
    paper height: 1440, 
    left  and right margin: 220, 
    top margin: 120,
    bottom margin: 100, column sep: 80, par skip: 4pt, par indent: 0.

\subsection{Eye Tracker Setup}
\label{sec:study-setup}

We recorded gaze with an EyeLink Portable Duo at $1000\,\mathrm{Hz}$. The passages were presented on a $27$-inch, $60\text{-}\mathrm{Hz}$ monitor at a resolution of $2560 \times 1440$ pixels, with physical display dimensions of $569\,\mathrm{mm}$ in width and $339\,\mathrm{mm}$ in height.
Participants sat approximately $50\,\mathrm{cm}$ from the display. We used the eye tracker's head-free remote mode and placed a target sticker on each participant's forehead to support head tracking. Before participants began the reading task, we performed a nine-point calibration and validation.

The double-column journal format requires the participant to focus on one column while the adjacent column may contain visual embellishments. 
A critical design challenge in this layout is determining the width of the central white gutter. If the gutter is too wide, the adjacent column is pushed too deep into the participant’s peripheral vision; the embellishments would suffer from peripheral crowding, a phenomenon where adjacent visual features in the periphery blend together into illegibility \cite{strasburger2011peripheral}. If the gutter is too narrow, the columns overlap in the foveal field, making it difficult to distinguish intent-driven gaze from accidental spillover.

The human visual field is limited and can be subjected to visual crowding, though exact estimates of the field vary. Existing optometric frameworks define the central visual field as extending up to $30^\circ$ eccentricity \cite{grosvenor2007primary}, whereas vision science research often locates the transition to the peripheral field at $10^\circ$ eccentricity \cite{loschky2019contributions}, restricting the central visual field to $<10^\circ$ \cite{strasburger2011peripheral}. 

\revadd{In reality, people read publications with different column layout at various screen size and resolution. To support generalizability, we adopted conservative visual-angle assumptions. 
In particular, when a participant is actively fixated on one column, the adjacent column will still reside within the boundary of central and peripheral vision.}
\revdel{We aimed to ensure, even by the most conservative visual angles, when a participant is actively fixated on one column, that the adjacent column resides within the boundary of central and peripheral vision.}
Therefore, we set the central column gutter to span exactly $10^\circ$ of total visual angle, positioning each column's inner margin $5^\circ$ of visual angle from the screen's vertical midline. 
At a $50\,\mathrm{cm}$ viewing distance, a $10^\circ$ visual angle corresponds to a physical screen width of approximately $8.75\,\mathrm{cm}$.
This translates to a gutter width of $405\,\mathrm{pixels}$ ($202\ \mathrm{pixels}$ on either side of the center).




\subsection{Participants}
\label{sec:participants}

Six participants completed the study and were included in the analysis. Their mean age was 24.7 years (range: 21--29); three identified as women and three as men. 
\revadd{All participants had graduate-level educational backgrounds and reported experience in reading academic papers.}
The average time spent on reading each passage is 125.19 seconds (SD=42.24).
Due to a technical failure with the eye-tracker, one trial (from P3, left-hand word-scale embellishment condition) resulted in corrupted data and was excluded from analysis.

\subsection{Study Procedure}
\label{sec:study-procedure}

The study interface first collected a participant identifier, age, and gender. 
After eye-tracker calibration and validation, participants read the four two-column passages while their gaze was recorded.
For the embellished passages, embellishments were placed in either the left or the right column only, and each participant saw embellishments in only one of the two columns.
Each participant viewed one embellished and one non-embellished passage in each of the data-heavy and data-light stimulus comparisons (see Table \ref{tab:stimuli}). 
We counter-balanced the presence of embellishment in left and right columns to be equivalent, and randomized the reading order of the four passages.


The following study design was repeated for all four passages. 
For each passage, participants first read the page and then opened the corresponding questionnaire block \revadd{after the passage is withdrawn}. 
Participants next answered three required multiple-choice comprehension questions about factual and relational information in the passage. 
Each comprehension question had three options.
The questions tested passage-specific details, such as reported quantities, comparisons between quantities, or the stated reason for an outcome. 
After the comprehension questions, participants rated six statements about their \revadd{subjective} reading experience on a seven-point scale from \textit{strongly disagree} to \textit{strongly agree} (\revdel{see} see Table \ref{tab:subjective}).

\section{Study Results}
\subsection{Saccade Analysis}
Re-reading behavior is often associated with disrupted reading \cite{navalpakkam2011using}.
To assess whether visual embellishments distract readers, we analyze two eye-movement measures associated with re-reading behavior: vertical saccades and cross-column saccades.
During normal English reading, saccades are predominantly horizontal. 
A higher proportion of vertical saccades therefore indicates greater deviation from the typical reading pattern. 
In double-column layouts, more frequent cross-column saccades indicate that readers switch between columns more often, suggesting increased re-reading behavior. 
Together, these measures \revdel{provide evidence} \revadd{are used as indicators} of disrupted reading behavior and visual distraction.

\begin{figure}
    \centering
    \includegraphics[width=\linewidth]{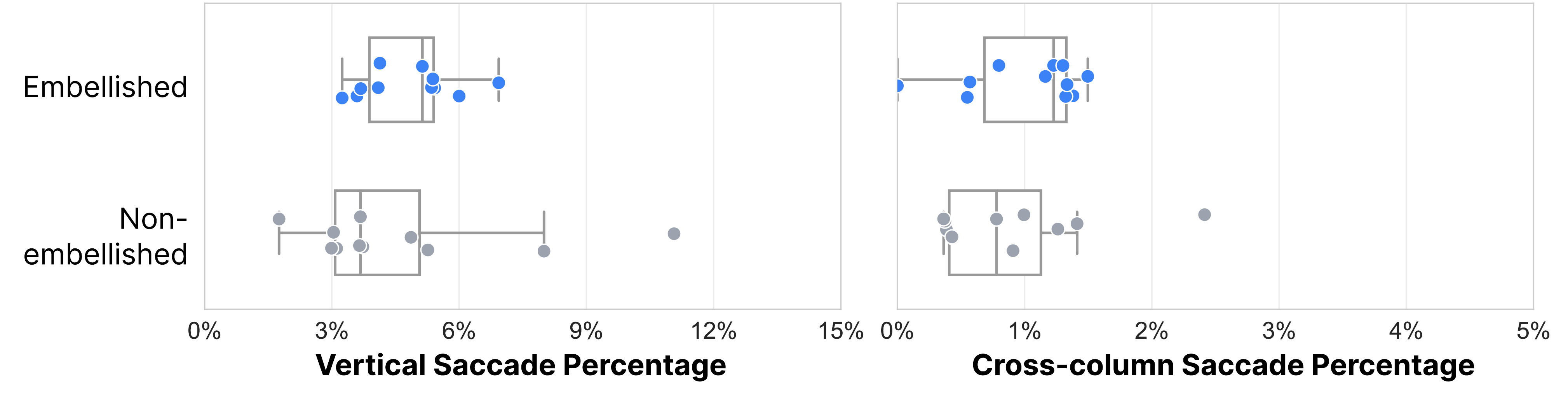}
    \vspace{-2em}
    \caption{\revdel{Embellishment conditions are related to slightly more vertical and cross-column saccades than non-embellishment conditions.}
    \revadd{There are small increases in vertical and cross-column average saccade rates for embellished conditions, with high variance}
    }
    \label{fig:vertical}
    \vspace{-1em}
\end{figure}

We observed a slight increase in the rates of both vertical (MD = 0.16\%) and cross-column saccades (MD = 0.13\%) in the embellished conditions relative to the non-embellished conditions, though with considerable variability across participants (Figure~\ref{fig:vertical}).
This variability likely stemmed from individual differences in top-down cognitive control, where readers with stronger attentional suppression are more likely to resist disruptions from peripheral salient cues, whereas others remain susceptible to salient distractors~\cite{fukuda2009human}.



\begin{figure*}[t]
    \centering
    \setlength{\tabcolsep}{0pt}

    \begin{tabular}{@{}
        >{\centering\arraybackslash}m{0.25\textwidth}
        >{\centering\arraybackslash}m{0.25\textwidth}
        >{\centering\arraybackslash}m{0.25\textwidth}
        >{\centering\arraybackslash}m{0.25\textwidth}
        @{}}

        \multicolumn{2}{c}{\small\bfseries Data-heavy}
        &
        \multicolumn{2}{c}{\small\bfseries Data-light}
        \\[3pt]

        \small\bfseries (a) Left embellished
        &
        \small\bfseries (b) Right embellished
        &
        \small\bfseries (c) Left embellished
        &
        \small\bfseries (d) Right embellished
        \\[2pt]

        \includegraphics[width=\linewidth]
            {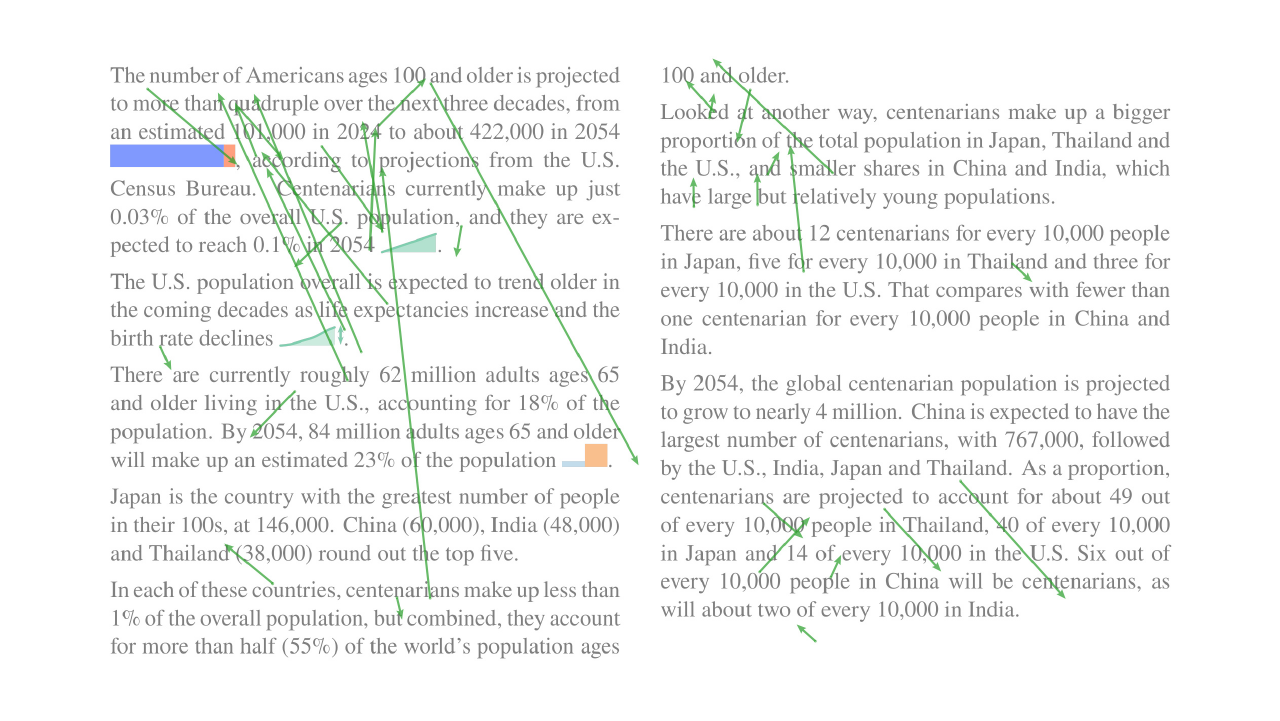}
        &
        \includegraphics[width=\linewidth]
            {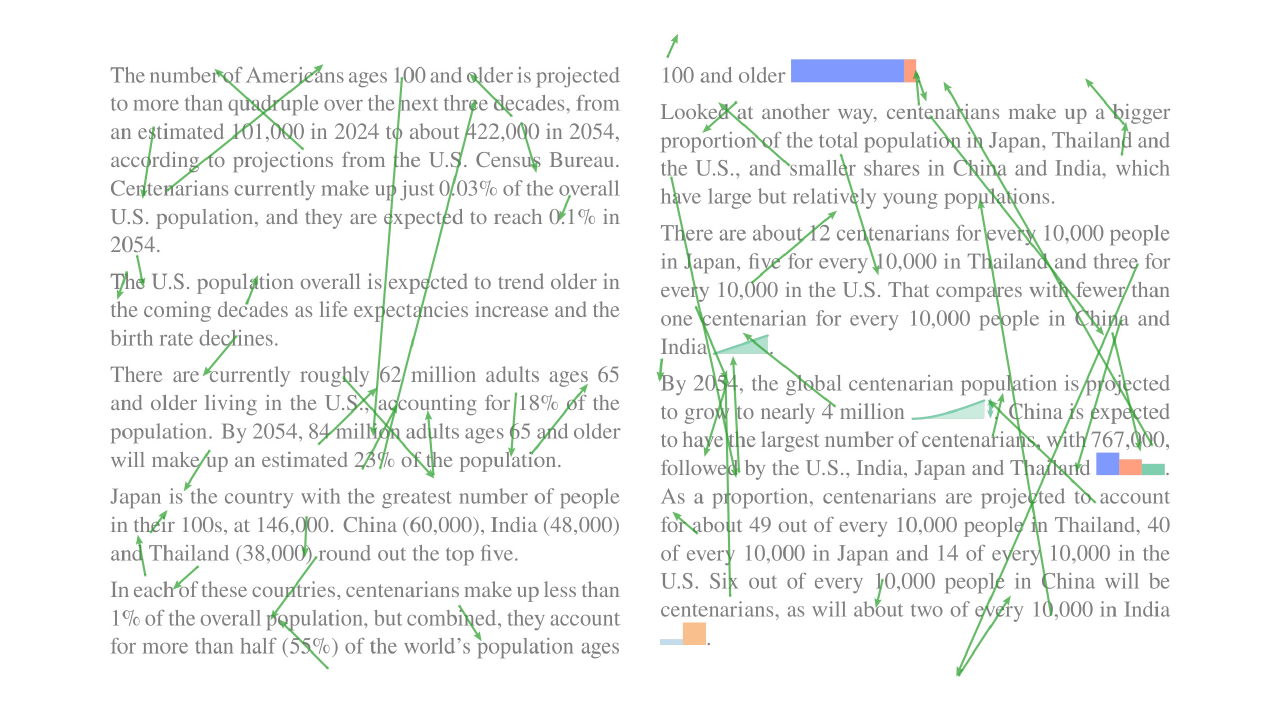}
        &
        \includegraphics[width=\linewidth]
            {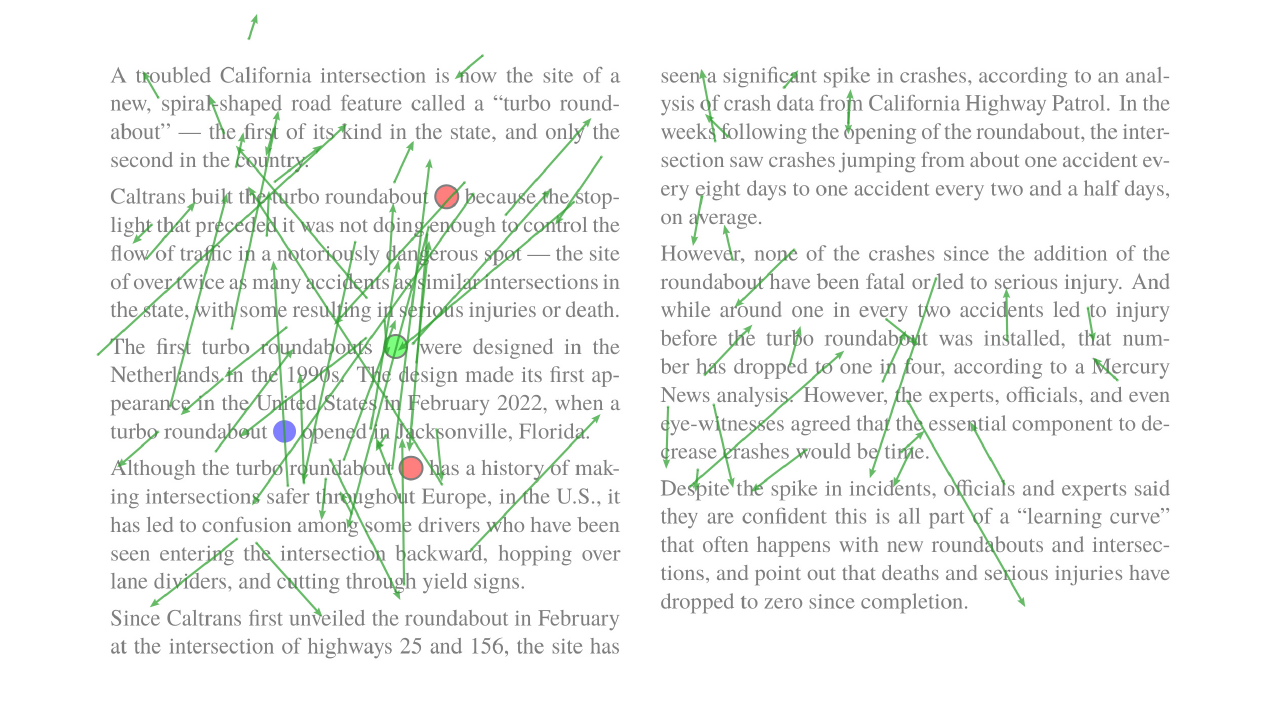}
        &
        \includegraphics[width=\linewidth]
            {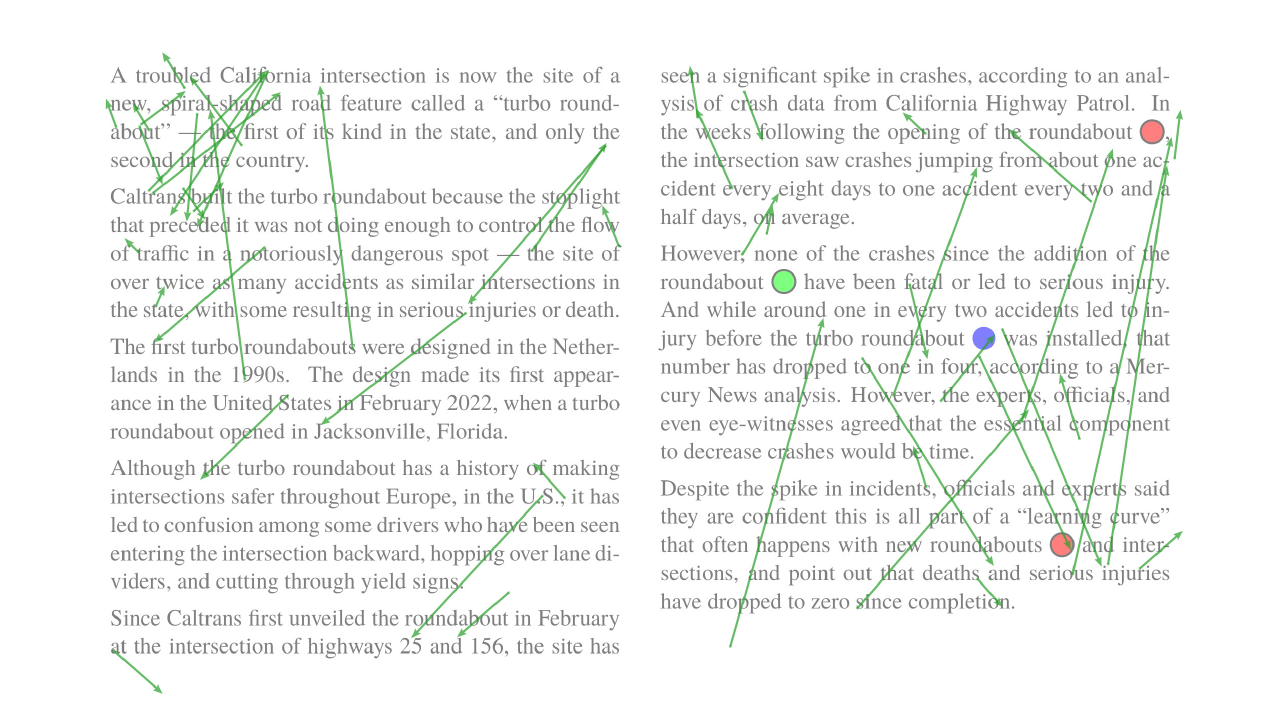}

    \end{tabular}
    \vspace{-2em}
    \caption{Aggregated gaze paths without horizontal and cross-column
    saccades. \revdel{More vertical saccades are observed around visual embellishments than in the corresponding non-embellished versions.}\revadd{Vertical-saccade concentrations appear near some embellishments. The strength of effect varies by four conditions, indicating variation by passage and placement.}}
    \label{fig:aggregated}
    \vspace{-2em}
\end{figure*}



\subsection{Gaze Path Patterns}

Participants exhibited distinct reading behaviors.
To ground these overarching trends in fine-grained eye-movement data, Figure~\ref{fig:p2-condition-comparison} presents the scan path of P2 as a representative case study that exemplifies the condition-level differences we observed. 
As shown in the figure, saccade density is higher around numbers and visual embellishments than other textual elements, forming localized saccade clusters.
\revdel{This suggests more prevelant re-reading behavior over those text regions, indicating that readers are spending more time reading numbers and visual embellishments, which is consistent with findings from GistVis~\cite{zou2025gistvis}.}\revadd{For P2, these localized clusters are consistent with re-reading around numbers and some embellishments, a pattern also observed in GistVis~\cite{zou2025gistvis}.} 


\revdel{We also observed that some participants re-read more than others.}\revadd{Using vertical-saccade rate as a descriptive proxy for re-reading, matched participant-level comparisons showed a higher average rate in embellished passages for three of six participants and a lower rate for three.} 
Figure~\ref{fig:aggregated} shows the aggregated scanpaths across  all six participants, contrasting embellished passages against their unembellished baselines.
We filter out horizontal saccades that are highly likely associated with normal reading behavior, as well as cross-column saccades, to focus on within-column re-reading behavior. 
\revdel{We observed that, for the same passage, within one column, there are more vertical saccades starting or ending around visual embellishments in the embellished passages compared to the non-embellished passages, despite the fact that data is missing for P3’s left-column word-scale embellishment.}\revadd{In several panels, vertical saccades start or end near embellishments more often in the embellished column; this pattern is less significant in the data-light right-embellished panel. As previously noted, data for P3's left-column word-scale trial are missing.}
\revdel{This suggests that visual embellishments disrupt fluent reading flow and break natural line progression, forcing readers to re-read text adjacent to the graphics, possibly explaining the longer reading time.}\revadd{The localized patterns motivate the hypothesis that some graphics interrupt line progression for some readers.}

\subsection{Subjective and Behavioral Data}
\noindent \textbf{Comprehension Score:}
After reading each passage, participants answered three \revadd{multiple-choice} comprehension questions. 
Comprehension was measured as the proportion of correctly answered questions. 
We aggregated comprehension scores by data density and visual embellishment (Table~\ref{tab:subjective}) and found no significant effects of visual embellishment in either data-heavy or data-light passages.




\noindent \textbf{Reading Experience:}
We  compared the reading experience across data intensity and visual embellishment.
Reading experience scores are derived by averaging the four reading experience \revadd{question }ratings (easy to understand, crowded, distracting, and easy to extract relevant info). 
Scores for crowded and distracting are reversed into \textit{not} crowded and \textit{not} distracting before calculation, so that higher scores consistently indicate a better reading experience.

Overall, readers reported a better experience with data-light passages than with data-heavy passages.
For data-light passages, readers preferred unembellished text over the embellished version, rating it as both less distracting and easier to understand \revadd{on average}.
For data-heavy passages, visual embellishments did not significantly improve or degrade the reading experience.

Notably, participants reported greater familiarity with the non-embellished passages in both data-light and data-heavy conditions. 
This might suggest that the increased fixations around visual embellishments do not inherently signify \revadd{distraction}\revdel{cognitive friction or reading difficulty}. 
It remains possible that readers re-examined embellishments due to novelty or deliberate attempts to encode numerical data. 
Disentangling these underlying motivations requires further investigation.

\begin{table}
\centering
\scriptsize
\setlength{\tabcolsep}{4pt}
\renewcommand{\arraystretch}{1.0}

\begin{tabularx}{\columnwidth}{
    @{}>{\raggedright\arraybackslash}X@{\hspace{4pt}}rcccc@{}
}
\toprule
\multicolumn{2}{c}{}
& \multicolumn{2}{c}{\textbf{Data-heavy}}
& \multicolumn{2}{c}{\textbf{Data-light}} \\
\cmidrule(lr){3-4}\cmidrule(l){5-6}

\multicolumn{2}{@{}l}{\textbf{Measurement}}
& \textbf{None}
& \textbf{Word-scale}
& \textbf{None}
& \textbf{Circles} \\
\midrule

\multicolumn{2}{@{}l}{Comprehension score (\%$\pm$SD)}
    & $83 \pm 18$
    & $78 \pm 34$
    & $89 \pm 17$
    & $94 \pm 14$ \\

\midrule

\multicolumn{2}{@{}l}{Familiar}
    & \highcell{54}{6.17}
    & \highcell{8}{4.33}
    & \highcell{63}{6.50}
    & \highcell{17}{4.67} \\

\multicolumn{2}{@{}l}{Interested}
    & \highcell{17}{4.67}
    & \highcell{25}{5.00}
    & \highcell{38}{5.50}
    & \highcell{17}{4.67} \\

\addlinespace[3pt]

\multicolumn{2}{@{}l}{Easy to understand}
    & \highcell{54}{6.17}
    & \highcell{29}{5.17}
    & \highcell{71}{6.83}
    & \highcell{33}{5.33} \\

\multicolumn{2}{@{}l}{\textit{Not} crowded}
    & \lowcell{21}{3.17}
    & \lowcell{13}{3.50}
    & \highcell{54}{6.17}
    & \highcell{38}{5.50} \\

\multicolumn{2}{@{}l}{\textit{Not} distracting}
    & \highcell{4}{4.17}
    & \lowcell{4}{3.83}
    & \highcell{54}{6.17}
    & \midcell{4.00} \\

\multicolumn{2}{@{}l}{Easy to extract relevant info}
    & \highcell{33}{5.33}
    & \highcell{8}{4.33}
    & \highcell{54}{6.17}
    & \highcell{38}{5.50} \\

\midrule

\multirow{2}{=}{\textbf{Reading experience}}
    & \textbf{Mean}
    & $\mathbf{4.71}$
    & $\mathbf{4.21}$
    & $\mathbf{6.33}$
    & $\mathbf{5.08}$ \\

    & SD
    & $1.39$
    & $1.03$
    & $0.44$
    & $1.11$ \\

\bottomrule
\end{tabularx}

\caption{Comprehension scores and subjective ratings by data intensity
and embellishment.
Reading experience questions use a 1--7 scale (n=6 per condition), where
1=Strongly Disagree and 7=Strongly Agree.
\revdel{Comprehension scores and reading experience are reported as
mean $\pm$ SD.}
}
\label{tab:subjective}
\vspace{-2em}
\end{table}


\noindent \textbf{Reading Speed:}
We  measured the reading speed for each passage and participant.
Reading time measured with the duration (ms) of eye gazing trials. 
For participant $i$ and passage $j$, $\mathrm{WPM}_{ij} = \text{word count}_j / (\text{duration}_{ij} / 60{,}000)$ and $\mathrm{AverageWPM}_i = \tfrac{1}{4}\sum_{j=1}^{4}\mathrm{WPM}_{ij}$.
We report a within-participant percentage difference, $100\times(\mathrm{WPM}_{ij}/\mathrm{AverageWPM}_i - 1)$, as shown in Table~\ref{tab:reading-speed}. 

\revdel{We observed that visual embellishments were associated with lower reading speed in data-light passages.}
\revadd{All six participants read the data-light circular-glyph passage more slowly than the data-light unembellished passage.}
In data-heavy passages, \revdel{however}, the effect differed across participants, with some readers exhibiting faster reading speeds and others slower \revdel{reading speeds}.


\begin{table}
\scriptsize
\setlength{\tabcolsep}{2.5pt}
\renewcommand{\arraystretch}{1.15}

\resizebox{\columnwidth}{!}{%
\begin{tabular}{
  lc
  *{4}{>{\centering\arraybackslash}m{1.45cm}}
}
\toprule
&
&
\multicolumn{2}{c}{\textbf{Data-heavy}} &
\multicolumn{2}{c}{\textbf{Data-light}} \\
\cmidrule(lr){3-4}\cmidrule(lr){5-6}

\textbf{ID}
&
\makecell{\textbf{Average WPM}}
&
\makecell{\bfseries None}
&
\makecell{\bfseries Word-scale}
&
\makecell{\bfseries None}
&
\makecell{\bfseries Circles} \\
\midrule

p1
& \cellcolor{wpmgray}216.6
& \cellcolor{wpmred!48}$-38.5\%$
& \cellcolor{wpmblue!52}$+41.2\%$
& \cellcolor{wpmblue!18}$+14.5\%$
& \cellcolor{wpmred!22}$-17.2\%$ \\

p2
& \cellcolor{wpmgray}266.9
& \cellcolor{wpmblue!32}$+25.4\%$
& \cellcolor{wpmred!32}$-25.6\%$
& \cellcolor{wpmblue!25}$+20.3\%$
& \cellcolor{wpmred!25}$-20.1\%$ \\

p3
& \cellcolor{wpmgray}114.5
& \cellcolor{wpmblue!10}$+8.1\%$
& \cellcolor{wpmblue!18}$+14.5\%$
& \cellcolor{wpmred!3}$-2.4\%$
& \cellcolor{wpmred!25}$-20.2\%$ \\

p4
& \cellcolor{wpmgray}157.3
& \cellcolor{wpmblue!5}$+4.3\%$
& \cellcolor{wpmred!12}$-9.2\%$
& \cellcolor{wpmblue!33}$+26.7\%$
& \cellcolor{wpmred!27}$-21.8\%$ \\

p5
& \cellcolor{wpmgray}159.8
& \cellcolor{wpmred!19}$-15.4\%$
& \cellcolor{wpmblue!15}$+11.7\%$
& \cellcolor{wpmblue!45}$+36.3\%$
& \cellcolor{wpmred!41}$-32.6\%$ \\

p6
& \cellcolor{wpmgray}172.3
& \cellcolor{wpmred!37}$-29.8\%$
& \cellcolor{wpmblue!71}{\color{white}$+56.4\%$}
& \cellcolor{wpmred!17}$-13.4\%$
& \cellcolor{wpmred!17}$-13.2\%$ \\

\bottomrule
\end{tabular}%
}
\caption{\revdel{Reading speed on data-light passages with visual embellishment, shown as deviation from each participant's average across the four passages}\revadd{Reading speed by participant and condition, shown as deviation from each participant's average across the four passages} (positive = faster, negative = slower)
}
\label{tab:reading-speed}
\vspace{-3em}
\end{table}

\section{Discussion}

\revdel{Our findings suggest that visual embellishments may alter reading dynamics by disrupting fluent line progression and triggering localized re-reading near embedded visuals and numerical data. 
Higher vertical saccade rates and increased fixation densities around the embellishments suggest increased reading time.
Together with subjective ratings for data-light passages, where readers favored unembellished text as easier to understand and less distracting, 
this may indicate that visual embellishments are associated with cognitive disruptions during reading, especially when they do not contain data-relevant information.}
\revadd{In this sample, embellished trials showed small descriptive increases in vertical and cross-column saccade rates and localized re-reading patterns for some participants. The data-light circular-glyph passage also received lower average `Easy to understand' and `Not distracting' ratings and was read more slowly than its unembellished comparison passage. These measures motivate the hypothesis that inline graphics can interrupt reading in some contexts.}

These results contrast with prior work demonstrating that word-scale visualizations of statistical `gists' reduce mental demand and subjective effort during reading \cite{zou2025gistvis}. 
This divergence may stem from key methodological differences; whereas GistVis evaluated interactive visualizations within a single-column layout, our study examined static embellishments in a two-column format. 
Furthermore, the substantial individual differences observed in our behavioral data suggest that visual embellishments do not impact all readers uniformly. 
Because increased gaze allocation can reflect aesthetic engagement or active data encoding rather than pure cognitive friction, our findings point to a complex interplay among document design, attentional control, and reader preferences. 
Ultimately, further research is required to delineate the trade-offs of inline word-scale visualizations and identify the precise contextual scenarios and user profiles for which they are most beneficial.


As word-scale visualizations become more popular in documents, researchers must take responsibility for evaluating their holistic impact. 
\revadd{Despite the limitations of the study, we still find preliminary signals that inline embellishments can influence both reading behavior and subjective reading experience.  This encourages future work with larger participant groups to identify the contextual and reader-specific factors that determine the effect of word-scale visualizations on reading. These include variations on passages, embellishment designs, and document layouts.}
\revdel{Future work should recruit a larger number of participants and consider other embellishment designs, passage topics, and layout modes, to establish robust design guidelines.}

\bibliographystyle{abbrv-doi}
\newpage
\bibliography{template}

@incollection{navalpakkam2011using,
  title={Using gaze patterns to study and predict reading struggles due to distraction},
  author={Navalpakkam, Vidhya and Rao, Justin and Slaney, Malcolm},
  booktitle={CHI'11 Extended Abstracts on Human Factors in Computing Systems},
  pages={1705--1710},
  year={2011}
}

@article{barach2021emojis,
  title={Are emojis processed like words?: Eye movements reveal the time course of semantic processing for emojified text},
  author={Barach, Eliza and Feldman, Laurie Beth and Sheridan, Heather},
  journal={Psychonomic Bulletin \& Review},
  pages={1--14},
  year={2021},
  publisher={Springer}
}

@inproceedings{lin2026four,
  title={A Four-Stage Framework of Visual Complexity and Trust as Mediated by Effort},
  author={Lin, Kylie and Guan, Hui and Rapp, David N and Bearfield, Cindy Xiong},
  booktitle={2026 IEEE 19th Pacific Visualization Conference (PacificVis)},
  pages={62--72},
  year={2026},
  organization={IEEE}
}

@article{loschky2019contributions,
  title={The contributions of central and peripheral vision to scene-gist recognition with a 180 visual field},
  author={Loschky, Lester C and Szaffarczyk, Sebastien and Beugnet, Clement and Young, Michael E and Boucart, Muriel},
  journal={Journal of Vision},
  volume={19},
  number={5},
  pages={15--15},
  year={2019},
  publisher={The Association for Research in Vision and Ophthalmology}
}

@article{strasburger2011peripheral,
  title={Peripheral vision and pattern recognition: A review},
  author={Strasburger, Hans and Rentschler, Ingo and J{\"u}ttner, Martin},
  journal={Journal of vision},
  volume={11},
  number={5},
  pages={13--13},
  year={2011},
  publisher={The Association for Research in Vision and Ophthalmology}
}

@inproceedings{zou2025gistvis,
  title={GistVis: Automatic generation of word-scale visualizations from data-rich documents},
  author={Zou, Ruishi and Tang, Yinqi and Chen, Jingzhu and Lu, Siyu and Lu, Yan and Yang, Yingfan and Ye, Chen},
  booktitle={Proceedings of the 2025 CHI Conference on Human Factors in Computing Systems},
  pages={1--18},
  year={2025}
}

@article{fukuda2009human,
  title={Human variation in overriding attentional capture},
  author={Fukuda, Keisuke and Vogel, Edward K},
  journal={Journal of Neuroscience},
  volume={29},
  number={27},
  pages={8726--8733},
  year={2009},
  publisher={Society for Neuroscience}
}

@article{goffin2014exploring,
  title={Exploring the placement and design of word-scale visualizations},
  author={Goffin, Pascal and Willett, Wesley and Fekete, Jean-Daniel and Isenberg, Petra},
  journal={IEEE Transactions on Visualization and Computer Graphics},
  volume={20},
  number={12},
  pages={2291--2300},
  year={2014},
  publisher={IEEE}
}

@inproceedings{lan2026evolving,
  title={The evolving duet of two modalities: a survey on integrating text and visualization for data communication},
  author={Lan, Xingyu and Li, Xi and Zhang, Yixing and Cheng, Mengqin and Wang, Jiazhe and Chen, Siming},
  booktitle={Proceedings of the 2026 CHI Conference on Human Factors in Computing Systems},
  pages={1--19},
  year={2026}
}

@article{beck2017word,
  title={Word-sized graphics for scientific texts},
  author={Beck, Fabian and Weiskopf, Daniel},
  journal={IEEE transactions on visualization and computer graphics},
  volume={23},
  number={6},
  pages={1576--1587},
  year={2017},
  publisher={IEEE}
}

@inproceedings{hoffswell2018augmenting,
  title={Augmenting code with in situ visualizations to aid program understanding},
  author={Hoffswell, Jane and Satyanarayan, Arvind and Heer, Jeffrey},
  booktitle={Proceedings of the 2018 CHI Conference on Human Factors in Computing Systems},
  pages={1--12},
  year={2018}
}

@inproceedings{brandes2013gestaltlines,
  title={Gestaltlines},
  author={Brandes, Ulrik and Nick, Bobo and Rockstroh, Brigitte and Steffen, Astrid},
  booktitle={Computer graphics forum},
  volume={32},
  number={3pt2},
  pages={171--180},
  year={2013},
  organization={Wiley Online Library}
}

@inproceedings{beck2016expert,
  title={An expert evaluation of word-sized visualizations for analyzing eye movement data},
  author={Beck, Fabian and Acurana, Yasett and Blascheck, Tanja and Netzel, Rudolf and Weiskopf, Daniel},
  booktitle={2016 IEEE Second Workshop on Eye Tracking and Visualization (ETVIS)},
  pages={50--54},
  year={2016},
  organization={IEEE}
}

@phdthesis{ceja2023limits,
  author    = {Ceja, Cristina R.},
  title     = {Limits for Binding Visual Information},
  school    = {Northwestern University},
  year      = {2023},
  type      = {Ph.D. dissertation},
  publisher = {ProQuest Dissertations \& Theses Global}
}

@book{grosvenor2007primary,
  title={Primary care optometry},
  author={Grosvenor, Theodore and Grosvenor, Theodore P},
  year={2007},
  publisher={Elsevier health sciences}
}

@article{carrasco2011visual,
  title={Visual attention: The past 25 years},
  author={Carrasco, Marisa},
  journal={Vision research},
  volume={51},
  number={13},
  pages={1484--1525},
  year={2011},
  publisher={Elsevier}
}

@article{franconeri2003moving,
  title={Moving and looming stimuli capture attention},
  author={Franconeri, Steven L and Simons, Daniel J},
  journal={Perception \& psychophysics},
  volume={65},
  number={7},
  pages={999--1010},
  year={2003},
  publisher={Springer}
}

@article{lavie2004load,
  title={Load theory of selective attention and cognitive control.},
  author={Lavie, Nilli and Hirst, Aleksandra and De Fockert, Jan W and Viding, Essi},
  journal={Journal of experimental psychology: General},
  volume={133},
  number={3},
  pages={339},
  year={2004},
  publisher={American Psychological Association}
}

@article{reingold2004saccadic,
  title={Saccadic inhibition in reading.},
  author={Reingold, Eyal M and Stampe, Dave M},
  journal={Journal of Experimental Psychology: Human perception and performance},
  volume={30},
  number={1},
  pages={194},
  year={2004},
  publisher={American Psychological Association}
}

@article{theeuwes2025attentional,
  title={Attentional capture and control},
  author={Theeuwes, Jan},
  journal={Annual Review of Psychology},
  volume={76},
  number={1},
  pages={251--273},
  year={2025},
  publisher={Annual Reviews}
}

@article{connor2004visual,
  title={Visual attention: bottom-up versus top-down},
  author={Connor, Charles E and Egeth, Howard E and Yantis, Steven},
  journal={Current biology},
  volume={14},
  number={19},
  pages={R850--R852},
  year={2004},
  publisher={Elsevier}
}

@article{bauer2010design,
  title={The design and evaluation of a graphical display for laboratory data},
  author={Bauer, David T and Guerlain, Stephanie and Brown, Patrick J},
  journal={Journal of the American Medical Informatics Association},
  volume={17},
  number={4},
  pages={416--424},
  year={2010},
  publisher={BMJ Group BMA House, Tavistock Square, London, WC1H 9JR}
}

@inproceedings{cohn2018emoji,
  title={Are emoji a poor substitute for words? Sentence processing with emoji substitutions.},
  author={Cohn, Neil and Roijackers, Tim and Schaap, Robin and Engelen, Jan},
  booktitle={CogSci},
  year={2018}
}

@inproceedings{goffin2015reading,
  title={Exploring the effect of word-scale visualizations on reading behavior},
  author={Goffin, Pascal and Willett, Wesley and Bezerianos, Anastasia and Isenberg, Petra},
  booktitle={Proceedings of the 33rd Annual ACM Conference Extended Abstracts on Human Factors in Computing Systems},
  pages={1827--1832},
  year={2015}
}

@article{goffin2016exploratory,
  title={An exploratory study of word-scale graphics in data-rich text documents},
  author={Goffin, Pascal and Boy, Jeremy and Willett, Wesley and Isenberg, Petra},
  journal={IEEE transactions on visualization and computer graphics},
  volume={23},
  number={10},
  pages={2275--2287},
  year={2016},
  publisher={IEEE}
}

@article{parsons2013testing,
  title={Testing the feasibility of small multiples of sparklines to display semimonthly income statement data},
  author={Parsons, Linda M and Tinkelman, Daniel},
  journal={International Journal of Accounting Information Systems},
  volume={14},
  number={1},
  pages={58--76},
  year={2013},
  publisher={Elsevier}
}

@book{tufte2006beautiful,
  author={Tufte, Edward R},
  title={Beautiful evidence},
  volume={1},
  year={2006},
  publisher={Graphics Press Cheshire, CT}
}

@book{wolf2018reader,
  title={Reader, come home: The reading brain in a digital world},
  author={Wolf, Maryanne},
  year={2018},
  publisher={Harper New York, NY}
}

@article{Mago2020BooksIT,
  title={Books in the Time of Screens: The Reading Habits of Slovenian Students},
  author={Jasna Mazgoň and M. Kovac and M. K. Sebart and Tadej Vidmar},
  journal={Universal Journal of Educational Research},
  year={2020},
  volume={8},
  pages={3916-3923}
}

@article{Miniukovich2017DesignGF,
  title={Design Guidelines for Web Readability},
  author={Aliaksei Miniukovich and A. D. Angeli and Simone Sulpizio and P. Venuti},
  journal={Proceedings of the 2017 Conference on Designing Interactive Systems},
  year={2017}
}

@article{Scaltritti2019InvestigatingEO,
  title={Investigating Effects of Typographic Variables on Webpage Reading Through Eye Movements},
  author={M. Scaltritti and Aliaksei Miniukovich and P. Venuti and R. Job and Antonella De Angeli and Simone Sulpizio},
  journal={Scientific Reports},
  year={2019},
  volume={9}
}

@article{roundabout,
  title={`{L}earning curve': California's first turbo roundabout near Bay Area is showing mixed safety results},
  author={Jillian D'onfro},
  journal={SF Gate},
  month={April},
  day=5,
  year=2024,
  url={https://www.sfgate.com/bayarea/article/turbo-roundabout-california-update-19404019.php
}
}

@article{smartphones,
  title={Take Cellphones Out of the Classroom, Educators Say},
  author={Tim Walker},
  journal={NEA Today},
  month={Oct},
  day=3,
  year=2024,
  url={https://www.nea.org/nea-today/all-news-articles/take-cellphones-out-classroom-educators-say
}
}

@article{centenarians,
  title={US centenarian population is projected to quadruple over the next 30 years},
  author={Schaeffer, Katherine},
  year={2024},
  journal={Pew Research Center},
  url={https://www.pewresearch.org/short-reads/2024/01/09/us-centenarian-population-is-projected-to-quadruple-over-the-next-30-years/}
}

@article{blackfriday,
  title={Online shopping has grown rapidly in US, but most sales are still in stores},
  author={DeSilver, Drew},
  year={2023},
  journal={Pew Research Center},
  url={https://www.pewresearch.org/short-reads/2023/11/22/online-shopping-has-grown-rapidly-in-u-s-but-most-sales-are-still-in-stores/}
}

@inproceedings{huth2024eye,
  title={Eye tracking on text reading with visual enhancements},
  author={Huth, Franziska and Koch, Maurice and Awad-Mohammed, Miriam and Weiskopf, Daniel and Kurzhals, Kuno},
  booktitle={Proceedings of the 2024 symposium on eye tracking research and applications},
  pages={1--7},
  year={2024}
}

@book{postman2005amusing,
  title={Amusing ourselves to death: Public discourse in the age of show business},
  author={Postman, Neil},
  year={2005},
  publisher={Penguin}
}

@article{naep2024,
title={The Nation’s Report Card: 2024 NAEP Results},
journal={
National Center for Education Statistics},
month={Jan},
day=29,
year=2025,
url={https://www.nationsreportcard.gov/reports/reading/2024/g4_8/}
}

@book{haidt2024anxious,
  title={The anxious generation: How the great rewiring of childhood is causing an epidemic of mental illness},
  author={Haidt, Jonathan},
  year={2024},
  publisher={Penguin}
}

@article{Hong2026ReviewAA,
  title={Review and Analysis of Scientific Paper Embellishments},
  author={Jiayi Hong and Yixuan Wang and Petra Isenberg and Ross Maciejewski},
  journal={ArXiv},
  year={2026},
  volume={abs/2603.20306},
  url={https://api.semanticscholar.org/CorpusID:286762733}
}

@article{Lu2026GraphingIU,
  title={Graphing Inline: Understanding Word-scale Graphics Use in Scientific Papers},
  author={Siyu Lu and Yanhan Liu and Shiyu Xu and Ruishi Zou and Chen Ye},
  journal={Proceedings of the Extended Abstracts of the 2026 CHI Conference on Human Factors in Computing Systems},
  year={2026}
}

@article{Perin2013SoccerStoriesAK,
  title={SoccerStories: A Kick-off for Visual Soccer Analysis},
  author={Charles Perin and Romain Vuillemot and Jean-Daniel Fekete},
  journal={IEEE Transactions on Visualization and Computer Graphics},
  year={2013},
  volume={19},
  pages={2506-2515},
  url={https://api.semanticscholar.org/CorpusID:165784}
}

@article{ayres2005split,
  title={The split-attention principle in multimedia learning},
  author={Ayres, Paul and Sweller, John},
  journal={The Cambridge handbook of multimedia learning},
  volume={2},
  pages={135--146},
  year={2005}
}

@inproceedings{quadri2024see,
  title={Do you see what i see? a qualitative study eliciting high-level visualization comprehension},
  author={Quadri, Ghulam Jilani and Wang, Arran Zeyu and Wang, Zhehao and Adorno, Jennifer and Rosen, Paul and Szafir, Danielle Albers},
  booktitle={Proceedings of the 2024 CHI conference on human factors in computing systems},
  pages={1--26},
  year={2024}
}

@article{fygenson2025cognitive,
  title={Cognitive affordances in visualization: Related constructs, design factors, and framework},
  author={Fygenson, Racquel and Padilla, Lace and Bertini, Enrico},
  journal={IEEE Transactions on Visualization and Computer Graphics},
  year={2025},
  publisher={IEEE}
}
\end{document}